\RequirePackage{fix-cm}

\documentclass[natbib,smallextended]{svjour3}       

\smartqed  
\usepackage{graphicx}
\usepackage{amssymb}

 \journalname{my journal}
\newcommand{\aap}{{Astron. Astrophys.}}

\def\dj{d\kern-0.4em\char"16\kern-0.1em}
\def\Dj{\mbox{\raise0.3ex\hbox{-}\kern-0.4em D}}

\begin{document}

\title{Particle acceleration in interstellar shocks
}


\author{Dejan Uro{\v s}evi{\' c}, Bojan Arbutina, Du{\v s}an Oni{\' c}}


\institute{D. Uro{\v s}evi{\' c}, B. Arbutina, D. Oni{\' c} \at
              Department of Astronomy, Faculty of Mathematics, University of Belgrade, Studentski trg 16, 11000 Belgrade, Serbia \\
              Tel: +381-11-2027827\\
              Fax: +381-11-2630151\\
              \email{dejanu@math.rs, arbo@math.rs, donic@math.rs}           
           }

\date{Received: date / Accepted: date}

\maketitle

\begin{abstract}

This review presents the fundamentals of the particle acceleration processes active in interstellar medium (ISM), which are essentially based on
the so-called Fermi mechanism theory. More specifically, the review presents here in more details the first order
Fermi acceleration process -- also known as diffusive shock acceleration (DSA) mechanism. In this case, acceleration is induced
by the interstellar (IS) shock waves. These IS shocks are mainly associated with emission nebulae (\hbox{H\,{\sc ii}} regions,
planetary nebulae and supernova remnants). Among all types of emission nebulae, the strongest shocks are associated with supernova remnants (SNRs).
Due to this fact they also provide the most efficient manner to accelerate ISM particles to become {high energy particles}, i.e.~cosmic-rays (CRs).
The review therefore focuses on the particle acceleration at the strong shock waves of supernova remnants.


\end{abstract}

\section{Introduction}
\label{intro}

Enrico Fermi suggested in his seminal paper (Fermi 1949) an elegant way of acceleration of charged particles to the energies of most Galactic cosmic-rays.
He derived that the energy gain in a collision between a particle and the magnetic field perturbation is $\propto (u/c)^2$. Here $u$ represents the speed of the
magnetic field perturbation, {and $c$ the speed} of high-energy particle. In the interstellar conditions, the acceleration process suggested by Fermi is
{ actually not very efficient, as $u/c$ is already a small value quantity, for non-relativistic shocks, which is then squared}. On the other hand, the fact that particles can gain energy in collisions with the moving magnetic field irregularities (so-called magnetic mirrors)
served as a fundamental starting point for all later acceleration theories seeking to explain CR creation in ISM. Approximately three decades later, this original
mechanism of particle acceleration became known as the so-called second order Fermi acceleration mechanism, because a new theory started developing -- the first order
Fermi acceleration or diffusive shock acceleration theory. Actually, in DSA acceleration model, the gain of particle energy through repeated shock crossings by successive head-on interaction with up and downstream disturbances,
is $\propto u/c$ (e.g.~Bell 1978a). In that sense, shock waves in ISM transform part of the bulk kinetic energy into thermal energy, {but also in} the CR acceleration.

As we will see, collisionless shock waves are crucial phenomena for particle acceleration in ISM. The formation of such a shock wave in magnetized IS plasma, the {charged
particle acceleration} and the magnetic field amplification are coupled processes. We can simply say that, on microscopic level, a shock is formed when the charged particles are reflected by an electromagnetic barrier. The ordinary collisions between particles are not {frequent enough  as in ordinary} hydrodynamical shocks (actual dissipation cannot be established by a simple
particle-particle interaction). However, we note that various processes, like resonant microinstabilities, which actually trigger colissionless shock-formation and evolution are still
not fully understood (see e.g.~Zekovi\' c 2019). The various shock waves in {ISM} are mainly associated with the known emission nebulae. There are generally three basic types of
emission nebulae: \hbox{H\,{\sc ii}} regions, planetary nebulae (PNe) and supernova remnants (SNRs). In \hbox{H\,{\sc ii}} regions around young, massive and hot stars, the existence
of shocks can be expected in weak (isothermal, radiative) form. Even though PNe are essentially \hbox{H\,{\sc ii}} regions, they have two associated shock waves: inner one which is strong
(adiabatic, non-radiative), and outer one which is weak (isothermal or radiative shock). Supernova remnants in {the} first two phases of evolution (free expansion and Sedov phases) have strong,
non-radiative shock waves. In the third, isothermal phase of evolution, these previously
strong shock waves rapidly lose kinetic energy and become weaker, radiative shocks. We expect particle acceleration at all kinds of collisionless shocks, strong {or weak}, but efficient
acceleration should be expected only {at} strong shocks of supernova remnants. In addition, the inferred Galactic cosmic-ray energy density (order of $1\ \mathrm{eV}\ \mathrm{cm^{-3}}$),
combined with the estimated time of the average CR existence in our galaxy, requires a Galactic CR production with a total power of about $10^{41}\ \mathrm{erg}\ \mathrm{s^{-1}}$
(Ginzburg \& Syrovatskij 1967). With the mean supernova (SN) explosion energy of around $10^{51}\ \mathrm{erg}$ and an inferred Galactic SN rate of several events per century, SNe and their
remnants are the most probable sources known to be able to provide such a power. {The} inner shocks of PNe are strong, but corresponding shock speeds are significantly lower than those of young SNRs.
Additionally PNe are short living objects with significantly lower energy contents. However, we can expect particle acceleration {in} PN shocks, but {at} this moment we do not have observational
evidence for creation of high energy particles in PNe -- this can be the subject of future research.

In this review, we present the {details of} DSA theory. In fact, we start with the basic particle acceleration model given by famous Enrico Fermi (Section 2.1),
and then continue with the elaboration of the so-called microscopic and macroscopic approaches to the test-particle DSA mechanism (Sections 2.2 and 2.3). In addition,
we discuss the consequences of the CR back-reaction through the non-linear DSA (Section 2.4). Finally, we emphasize the most important observational signatures which
confirm DSA mechanism in SNRs (Section 3).

\section{Fermi acceleration}

\label{sec:1}

\subsection{The original Fermi approach}

In this subsection we present {the original} Fermi (1949) approach. Let us assume a moving magnetic perturbation of ISM (e.g.~a magnetized cloud) along the $x$-axis, moving with speed $u$
in the fixed reference frame of the observer. Let us also consider a particle with speed $v$ also in the reference frame of observer; $u$ and $v$ are {speeds}
of a moving magnetic perturbation and a particle, respectively, before collision. In the beginning, we consider the case where {the test-particle} moves in a direction opposite to the motion of magnetized cloud, and has initial energy $\mathcal{E} = m c^2 +E$, where $\mathcal{E}$ is the total energy,
$m$ is the particle rest mass, $c$ is the speed of light, and $E$ is the particle kinetic energy. We assume that particle is {gyrating} in a low-density medium with a turbulent magnetic field.
When it collides {with the} cloud, a charged particle is reflected by {the} so-called magnetic mirror and comes back with the same pitch angle $\theta$ and the direction of motion of its
guiding center being inverted. Applying the relativistic transformations between a 'laboratory' (observer's) reference frame at rest and the moving (primed) reference frame of the cloud,
the energy and the momentum of the particle before collision are respectively: \begin{equation}
\mathcal{E}'=\gamma_u(\mathcal{E}+up_x),\quad
p_x'=\gamma_u(p_x+u\mathcal{E}/c^2), \qquad \gamma_u
=1/{\sqrt{1-u^2/c^2}}.
\end{equation}

\noindent The collision between particle {and a magnetized cloud} is elastic, and in 'primed' reference frame the total energy of the test charged particle does not change after collision
$(\mathcal{E}'_{\rm before\, coll.}=\mathcal{E}'_{\rm after\, coll.})$. Also, the intensity of the particle momentum stays constant after collision in 'primed' reference frame,
but with opposite sign, {i.e.} $p_x'$ will be transformed to $-p_x'$ after collision. We can {move} back to the observer's reference frame by using similar reasoning,
as $u$ and $v$ vectors are oriented in the same direction after collision ($\mathcal{E}'=\gamma_u(\mathcal{E}-up_x),\, p_x'=\gamma_u(u\mathcal{E}/c^2-p_x)$). {We} change
$p_x'$ into $-p_x'$ to account for the inversion of the direction of propagation of the particle, and {obtain}: \begin{equation}
\label{eq36}
\mathcal{E_{\rm after\, coll.}}= \gamma_u(\mathcal{E}'+up_x') = \gamma_u^2(\mathcal{E}+2up_x+u^2\mathcal{E}/c^2),
\end{equation}

\noindent i.e., since (by using $p=\gamma mv$ and $\mathcal{E}=\gamma mc^2$), $p_x/\mathcal{E}=p{\cos}\theta/\mathcal{E}=v\cos\theta/c^2$, \begin{equation}
\label{eq37}
\frac{\mathcal{E_{\rm after\, coll.}}}{\mathcal{E}}=\frac{1+2(u/c)(v/c)\cos\theta+(u/c)^2}{1-(u/c)^2}
\approx 1+2(u/c)(v/c)\cos\theta+2(u/c)^2.
\end{equation}

\noindent The resulting expression in Eq.~(3) is obtained under particular approximation
$(1-(u/c)^2)^{-1} \approx 1+(u/c)^2$ and by neglecting all terms in which $u/c$ {appears with the
power} $\ge3$.

The probabilities of head-on and head-tail collisions are proportional to the intensity of relative velocities of approach of the
particle and the cloud, i.e.~$v+u \cos \theta$ and $v - u\cos\theta$ in the direction of the test-particle trajectory, respectively (for $\cos \theta > 0$).
Assuming high energy test-particle, for $v\approx c $ and $0<\theta<\pi$ we can write this probability as proportional
to $1+(u/c)\cos \theta$, and average the second term in Eq.~(3) to obtain (Longair 2011): \begin{equation} \label{eq38}
\left\langle \frac{2u}{c}\cos \theta \right\rangle = \frac{2u}{c} \frac{\int
_{-1} ^1 \mu (1+(u/c)\mu) d\mu}{\int _{-1} ^1 (1+(u/c)\mu) d\mu}=
\frac{2}{3} \Big(\frac{u}{c}\Big)^2, \ \ \ \mu = \cos \theta,
\end{equation}
i.e. \begin{equation} \label{eq39}
\langle {\Delta E}/{E} \rangle = \frac{8}{3}
\Big(\frac{u}{c}\Big)^2,
\end{equation} \noindent where we present the expression for the particle kinetic energy $E$ increment, under assumption of {negligible} rest energy in comparison to total energy of accelerated particles.

Under the starting assumption that a high energy test-particle has the same direction as a moving magnetic
perturbation (head-tail, rear-on or overtaking collision), transformations between 'primed' and observer's reference frames
for the total energy and momentum of the particle are {inverted} ({with respect to the} cases before and after collision).
In the case before collision for the transformation into the 'primed' frame we use $\mathcal{E}'=\gamma_u(\mathcal{E}-up_x),\, p_x'=\gamma_u(u\mathcal{E}/c^2-p_x)$.
In the case after collision for the return into the observer's reference frame, we have: \begin{equation}
\mathcal{E_{\rm after\, coll.}}= \gamma_u(\mathcal{E}'+up_x')=\gamma_u^2(\mathcal{E}-2up_x+u^2\mathcal{E}/c^2),
\end{equation}

\noindent Due to this, the analogue expression to Eq.~(3), for the case of overtaking collision is: \begin{equation} \label{eq37}
\frac{\mathcal{E_{\rm after\, coll.}}}{\mathcal{E}}\approx 1-2(u/c)(v/c)\cos\theta+2(u/c)^2,
\end{equation}

\noindent which after the same procedure applied previously for head-on collision and assuming that the probability of head-tail collision is
proportional to $1-(u/c)\cos \theta$ gives again $\langle {\Delta E}/{E} \rangle = 8/3(u/c)^2$.

As we can see from previously given equations, in the original version of the Fermi acceleration theory (Fermi 1949) $+$ and $-$ signs in front of the linear term $(u/c)$ correspond,
respectively, to the head-on (approaching) or head-tail (receding) motion of the magnetic mirror, and hence to a gain or a loss of energy. As a result, this linear term disappears
from the corresponding equations. Since the direct collisions (head-on) are statistically more numerous than the overtaking ones (for the same reason that car windscreens get wetter
than rear windows), there is a net energy gain for the charged particle, whose energy increases continuously due to the numerous accumulated collisions,
and this gain is represented by the second order dependence $(u/c)^2$ that survived in Eq.~(5).

\subsection{DSA - microscopic approach}

In the interstellar conditions, the original mechanism proposed by Enrico Fermi is not very efficient and we should try to establish a model in which only head-on collisions exist.
It means that the linear term $u/c$ (Eqs.~3 and 7) does not vanish. A modern version of the first order Fermi acceleration (DSA theory) was developed
independently by Axford et al.~(1977), Krymsky (1977), Bell (1978a) and Blandford \& Ostriker (1978). There {were two approaches} to the problem: macroscopic
(e.g.~Blandford \& Ostriker 1978) and microscopic (Bell 1978a). In the rest of this subsection we will follow the derivation by Bell (1978a).

\begin{figure}[h]
  \includegraphics[bb=0 0 2800 1600,width=120mm,keepaspectratio]{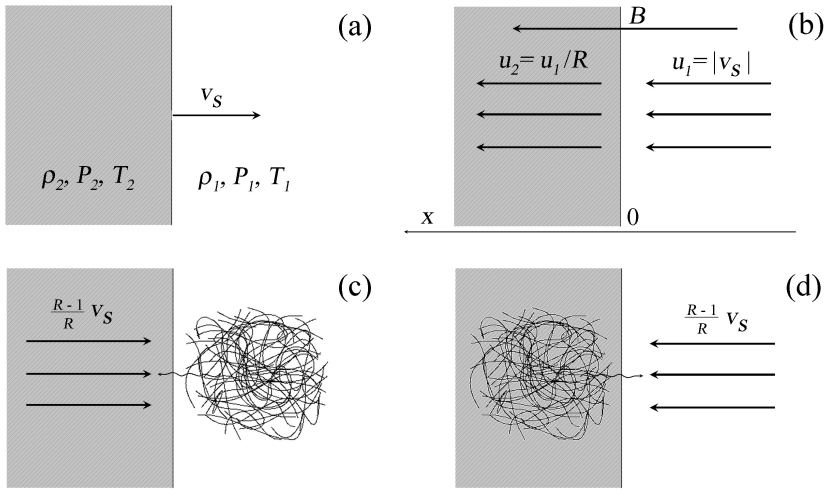}
\caption{Diffuse shock acceleration of high energy particles in the vicinity of a strong shock wave. Adapted from
  Longair (2011).}
\end{figure}

Let us start with consideration of a strong shock wave moving through the surrounding medium with speed $v_{\rm s}$ in the observer's (rest) frame (Fig.~1a). In the moving frame
which is connected to the shock (Fig.~1b), the upstream gas flows through the shock front with speed $u_1 = |v_{\rm s}|$. We assume that the shock surface is orthogonal
to the magnetic field lines $B$ (i.e.~parallel MHD shock wave). The equation of continuity (mass conservation) requires $\rho _1 u_1 = \rho_2
u_2$ and {the} so-called Rankine-Hugoniot relations give us $\rho _2/\rho _1 =
u_1/u_2 = R = (\gamma +1)/(\gamma -1)$, where $\gamma$ is the
ratio of specific heats of the gas. The compression ratio $R$ equals $4$ in the case of the strongest shocks.

Let us assume the existence of high-energy particles in {the} upstream region (ahead of the shock front) whose distribution function is isotropic in the frame of reference
in which the gas is at rest (Fig.~1c). This reference frame is equivalent to the observer's (rest) frame. These so-called test-particles cross the
shock front and so encounter gas behind the shock traveling at {the speed} $3/4v_{\rm s}$ if the shock is very strong. They are scattered by the magnetic irregularities
in the downstream region (behind the shock front) and they receive a small amount of energy $\Delta E/E \sim v_{\rm s}/c$. Their distribution function is also isotropized {-- their velocity distribution is now isotropic in the reference frame in which the downstream gas is at rest,} and they
are ready to move back to the upstream region. Now, we consider the opposite process when high-energy test-particle recross the shock front, from the downstream to the upstream region
(Fig.~1d). Due to that, when test
particles cross the shock front from downstream to upstream, they encounter the upstream gas moving at speed $3/4v_{\rm {\rm s}}$ and are again able to receive a small amount of energy.
Actually, in the upstream region they are scattered back by the Alfv\'en waves excited by the energetic particles themselves, as they move at super-Alfv\'{e}nic velocities
and attempt to escape from the shock. As a result, particles can recross the shock front again. In each shock recrossing, after collisions with downstream and upstream magnetic mirrors, they receive a small amount of energy -- there are no collisions
in which a particle {loses} energy.

After this more qualitative explanation, we move on to a more detailed theory of DSA (see Bell 1978a, Lequeux 2005, Longair 2011, {Arbutina 2017}). We consider the existence of test-particles 
and introduce their so-called phase-space distribution function, in particular its isotropic part $f(x,p,t)$. These charged particles are injected into the planar, {one-dimensional MHD shock,
parallel or nearly parallel (with shock velocity directed along the $x$-axis practically aligned with mean (galactic) magnetic field)} with speed much higher than that of the shock, $v \gg v_{\rm s}$. In these collisionless shocks, the average, homogeneous magnetic field plays essentially no role, while fluctuations
in that average field play an important role as scattering centers in the downstream region. In fact, we consider collisionless shock waves as ordinary hydrodynamic shocks that propagate
through the homogeneous magnetized plasma, as well as additional scattering centers. We choose a reference frame in which the shock is stationary (Fig.~1b). The gas flows from upstream,
$(x<0)$ with speed $u_1 = |v_{\rm s}|$, to downstream where $u_2 < |v_{\rm s}|$. In that sense, we are dealing with diffusion in a fluid moving with different velocities on either side of the shock.
Diffusion in a fluid is introduced by diffusion-advection equation near a discontinuity (Drury 1983, Blandford \& Eichler 1987), {which represents general kinetic equation for
cosmic-rays (Skilling 1971, 1975), sometimes called Parker's transport equation because it was derived for the first time in Parker (1965)}: \begin{equation}
 \label{eq40}
\frac{\partial{f}}{\partial{t}}+u\frac{\partial{f}}{\partial{x}}=\frac{\partial}{\partial{x}}{\bigg[}D(x,p)\frac{\partial{f}}{\partial{x}}{\bigg]}+\frac{1}{3}\frac{\partial{u}}{\partial{x}}p\frac{\partial{f}}{\partial{p}}
+Q(x,p),
\end{equation}

\noindent where \begin{equation}
D(x,p)=\frac{\lambda v}{3}
\end{equation}

\noindent is the diffusion coefficient in the $x$ direction for particles
with speed $v$. The mean free path of particles ($\lambda$) is assumed to be $\lambda {\sim} r_g$, \begin{equation}
\label{eq42}
r_{\rm g}=\frac{p_\perp}{eB}\frac{A}{Z},
\end{equation}

\noindent where $r_{\rm g}$ is the gyro-radius of charged particle with the {atomic mass number $A$, the charge number $Z$
in the magnetic field $B$.} Here the elementary charge is given by the symbol $e$ and the perpendicular
component of momentum $p_\perp$. The last term in Eq.~(\ref{eq40}) describes injection and we assume
$\delta$-function for injection at the shock.

{Equations (9) and (10) actually assume standard isotropic Bohm diffusion. It is important to note that the Bohm limit for the parallel diffusion coefficient was derived in Shalchi (2009)
and tested numerically by Hussein \& Shalchi (2014).}

Under the assumption of stationarity we have ${\partial{f}}/{\partial{t}}=0$ and
${\partial{u}}/{\partial{t}}=0$ outside the shock. Additionally we assume constant
downstream velocity (${\partial{u}}/{\partial{x}} =0$). Due to these simplifications, the diffusion-advection equation for particles which enter the downstream region 
(Eq.~(8)) becomes: 
\begin{equation}
u_2\frac{\partial{f}}{\partial{x}}-
\frac{\partial}{\partial{x}}\left[D(x,p)\frac{\partial{f}}{\partial{x}}\right]=0.
\end{equation}

\noindent The general solution of this partial differential equation, obtained by double integration over $x$,
is: \begin{equation}
f(x,p)=A(p)+B(p)\exp\left(\int{\frac{u_2}{D(x',p)}dx'}\right),
\end{equation}

\noindent where $A$ and $B$ are the integration constants. The diffusion coefficient
$D(x',p)$ takes finite values, while $x$ extends to infinity. Due to this the second term diverges,
unless $B=0$. A physical solution therefore requires that $f(x,p)=A(p)$ i.e.~it is {a function of momentum only}. The flow of particles far away from the shock
into the downstream region, i.e.~the current density of particles that escape from
the shock far downstream is $u_2n(x,p)-D(x,p)(\partial{n}/\partial{x})$, which is equal to $u_2
n(0,p)$, where the number density of particles $n = \int _0^\infty 4\pi p^2 f(p) dp$, is defined
for isotropic distribution. The flux density, i.e.~the rate at which particles are crossing and
recrossing the shock is: \begin{eqnarray}
\int\int f v_x p'^2 dp' d\Omega && = \int
_0 ^\infty  2\pi v(p') f(p') p'^2 dp' \int _0 ^{\pi/2} \cos \theta
\sin \theta d\theta=
\\
&&= \frac{1}{2} v n \int _0^1 \mu d\mu =
\frac{1}{4} v n(0,p),
\end{eqnarray}

\noindent where $d\Omega$ is the elementary solid angle, $\mu = \cos \theta$ and {we assume} $f(p')=\frac{n}{4\pi p^2} \delta (p-p')$. {We use the} well-known behavior
of the $\delta$-function $f(x)=\int f(x')\delta(x'-x)dx'$. Finally, we obtain the escape probability in the form:
\begin{equation}
\eta= \frac{u_2 n(0,p)}{\frac{1}{4}v n(0,p)} = 4\frac{u_2}{v}.
\end{equation}

\noindent For {relativistic particles} ($v\approx c$) and non-relativistic shock waves, this probability is rather low.

Now we focus to the reference frames in which DSA mechanism can be analytically described. We can define a rest frame attached to the magnetic mirror which is in the downstream region
of a shock wave. This magnetic mirror has velocity $u_1-u_2$ relative to the reference frame of upstream isotropically {distributed} particles (see Fig.~1c). On the other hand, in the upstream
region we can define another magnetic mirror with the same velocity $u_1-u_2$ (this is velocity in the reference frame connected to downstream isotropically {distributed} particles (Fig.~1d)).
The upstream magnetic mirror is {at} rest in the upstream fluid (observer's) reference frame. Due to this we define another rest reference frame to the upstream magnetic mirror.
These two rest reference frames tied to both magnetic mirrors are necessary for the next derivation. When a test-particle with injection energy $E_0$ in the fixed, upstream fluid (observer's)
reference frame goes from upstream (region 1) to downstream (region 2), its energy (according to Eq.~(\ref{eq36}), ${E}'=\gamma_u({E}+up_x)$) is: \begin{equation}
\label{eq46} E_1^{\rm down.\, mirror}\approx
E_0{\bigg[}1+(u_1-u_2)\frac{v_{11}\cos{\theta_{11}}}{c^2}{\bigg]},
\end{equation}

\noindent with $\gamma_u\approx1$ (because of $(u_1-u_2)^2/c^2\ll 1$) for non-relativistic shocks; here $E_1^{\rm down.\, mirror}$ represents the test-particle energy in the rest reference
frame of downstream magnetic mirror. When the same particle comes back to region 1, its energy in the rest frame of the upstream magnetic mirror is: \begin{equation}
\label{eq47} \frac{E_{1}^{\rm up. \, mirror}}{E_0}\approx
{\bigg[}1+(u_1-u_2)\frac{v_{11}\cos{\theta_{11}}}{c^2}{\bigg]}{\bigg[}1+(u_1-u_2)\frac{v_{12}\cos{\theta_{12}}}{c^2}{\bigg]}.
\end{equation}

\noindent where first index in $v$ and $\theta$ denotes cycle number, while second index denotes passing from region 1 to region 2 (index 1) and from region 2 to region 1 (index 2).
Eq.~(17) represents a starting expression for the acceleration of test-particle by the DSA mechanism and it is analogue to Eq.~(3). Here, $v_{12}$ and $\theta_{12}$ are {assumed} to be in the upstream mirror reference frame as it is
presented by Bell (1978a) -- we do not use these quantities in the downstream mirror reference frame (as {in} Lequeux 2005). For averaging per angle $\theta$, the necessary condition
is that angles $\theta_{11}$ and $\theta_{12}$ should be expressed in the same reference frame (we {chose it to be the} upstream mirror frame). Eq.~(17) has different form in Bell (1978a) -- the
reason for that is opposite direction of measuring $\theta$. We assume here the direction as given in Lequeux (2005). The sign $\approx$ instead equality in Eq.~(17) is a result of
transformation from the downstream to the upstream mirror frame ($p_x^{\rm up. \, mirror}=\gamma_{u_1-u_2}(p_x^{\rm down.\, mirror}+(u_1-u_2)E_1^{\rm down.\, mirror}/c^2$), and by using
approximation $\gamma_{u_1-u_2}=(1-((u_1-u_2)/c)^2)^{-1/2} \approx 1+1/2((u_1-u_2)/c)^2$, where finally we have term $v_{12}\cos\theta_{12}/c^2$.

We emphasize again that our test-particle is {already} highly energized ($v \approx c$) at the injection. Due to this, by using $v_{k1}/c\approx1$ and $v_{k2}/c\approx1$ which provide that
the test-particle gets the same portion of energy in every interaction with magnetic mirrors, the energy of test-particle after $l$ cycles is: \begin{equation}
\frac{E_l}{E_0}=\prod_{k=1}^{l}{\frac{E_{k}^{\rm up.\, mirror}}{E_0}}=\left(\frac{E_1^{\rm up. \, mirror}}{E_0}\right)^l,
\end{equation} \noindent where \begin{equation}
{\frac{E_{k}^{\rm up.\, mirror}}{E_0}}=\frac{1}{E_0}{\bigg[}1+(u_1-u_2)\frac{\cos{\theta_{k1}}}{c}{\bigg]}{\bigg[}1+(u_1-u_2)\frac{\cos{\theta_{k2}}}{c}{\bigg]},
\end{equation} \noindent i.e.~\begin{equation}
\ln\left({\frac{E_l}{E_0}}\right)=l\ln\left(\frac{E_1^{\rm up. \, mirror}}{E_0}\right).
\end{equation}

\noindent For a significant energy increase, $l$ must be at least of the
order of $c/(u_1-u_2)$. The distribution of $\ln
({{E_l}/{E_0}})$ will be strongly concentrated around the mean, so we can treat all particles completing
$l$ cycles as having their energy increased by the same amount, and by using Eq.~(19) with ${\frac{E_{k}^{\rm up.\, mirror}}{E_0}}=\frac{E_1^{\rm up. \, mirror}}{E_0}$, we obtain: \begin{equation}
\ln\left(\frac{E_l}{E_0}\right)=l\Big[
\Big\langle\ln\big(1+\frac{u_1-u_2}{c}\cos\theta_{k1}\big)\Big\rangle+\Big\langle\ln\big(1+\frac{u_1-u_2}{c}\cos\theta_{k2}\big)\Big\rangle
\Big] .
\end{equation} \noindent After the expansion of logarithm function in power series we find: \begin{equation}
\ln\left(\frac{E_l}{E_0}\right)\approx l \frac{u_1-u_2}{c}\left[\langle\cos\theta_{k1}\rangle + \langle\cos\theta_{k2}\rangle\right] .
\end{equation} \noindent The number of particles that cross the shock between angles
$\theta$ and $\theta +d\theta$ is proportional to
$2\pi{\sin\theta}{\cos\theta}d\theta$. After averaging over angles
from 0 to $\pi/2$ we have: \begin{equation}
\langle \cos\theta_{k1}\rangle = \langle \cos\theta_{k2}\rangle =
\frac{ \int _0^1 \mu ^2 d\mu }{ \int _0^1 \mu d\mu}= \frac{2}{3}.
\end{equation}\noindent Finally, \begin{equation}
\label{eq53}
\ln\left(\frac{E_l}{E_0}\right)=\frac{4}{3}l\frac{u_1-u_2}{c}.
\end{equation}

The probability  of completing at least $l$ cycles and reaching
energy $E\geq E_l$ is given by: \begin{equation}
\label{eq54}
{P}_l = \frac{N(E\geq E_l)}{N_o} = \zeta ^l, \ \ \ \zeta = 1-\eta,
\end{equation}

\noindent where $\zeta$ is the probability of staying in the acceleration
process after one cycle and $N_0$ is the initial number of particles. Each particle has the same energy $E_0$ at the injection.
The number of particles that reached $E\geq E_l$ is {designated} by $N(E\geq E_l)$. Using Eqs.~(15) and (25) leads to:
\begin{equation}
\label{eq55} \ln {P}_l = l \ln \Big( 1 - \frac{4u_2}{c}\Big)
\approx - l \frac{4u_2}{c} = - \frac{3u_2}{u_1-u_2}
\ln\left(\frac{E_l}{E_0}\right).
\end{equation}

\noindent By combining Eqs.~(\ref{eq53}) and (\ref{eq55}) we have: \begin{equation}
\label{eq56} {N(E\geq E_l)} =
{N_o}\left(\frac{E_l}{E_0}\right)^{1-\mathnormal{\Gamma} } = \int _{E_l}^\infty
N(E) dE , \ \ \ \mathnormal{\Gamma} = \frac{R+2}{R-1}, \ \ \ R=\frac{u_1}{u_2}.
\end{equation}

\noindent At the end, the differential particle energy spectrum has a form: \begin{equation}
\label{eq57} {N(E) } = K E^{-\mathnormal{\Gamma} } ,
\end{equation}

\noindent where $K$ is constant. In the case of strong shocks, the compression $R=4$, so
the energy index $\mathnormal{\Gamma}=2$. We emphasize here that Eq.~(28) is derived {by using
the assumption} that approximately all $N_0$ particles at the beginning of the acceleration process become
cosmic-rays. It means that all of them reach energy $E_l$. The main reason for the validity of this
assumption is based on the starting assumption that all $N_0$ particles are highly energized at the injection
i.e.~$v/c\approx1$, which provides $\zeta\approx1$. Due to this, in the limiting situation the integration in
Eq.~(27) can be from $E_0$ to $\infty$.

As a summary of the microscopic approach of test-particle DSA presented in this subsection, we emphasize that the net energy gain in the acceleration process {is} substantially
higher if there {are} only direct (head-on) collisions. It provides that energy increase is $\Delta E/E \propto (u_1-u_2)/c$ in each collision. As we mentioned earlier, the microscopic
DSA model is based on multiple transition of one charged particle through the shock discontinuity from upstream to downstream region and vice-versa. In every passage (head-on) across
the shock, independent from which side of the shock the passage occurs, the test-particle gains energy. Owing to this DSA is a more efficient process than original Fermi 2 mechanism (Eq.~(5)),
where the overtaken collisions exist. {If an astrophysical source is linked to a shock wave, we can generally expect acceleration of charged particles from the medium around the shock.
Particles such as protons and other heavier ions can be accelerated very efficiently to $\sim10^{15}$ eV by DSA process ({see e.g. Bell et al. 2013 and references therein}). Of course, electrons can also be accelerated to the ultra-relativistic
energies ($\sim10^{12}$ eV) at the strong shocks of SNRs, {but will suffer more significant energy losses (see e.g. Blasi 2010)}. Once again, we emphasize that the basic assumption of this derivation is that before
the actual start of DSA mechanism the particle has very high velocity $v\gg u$. Finally, this theory predicts a power-law energy spectrum of accelerated particles (Eq.~(28)). As we showed here, this theory
also predicts a value of $\mathnormal{\Gamma}=2$ for accelerated particles at the strong shock waves with the compression ratio $R=u_1/u_2=4$.} DSA model can be generalized to include lower starting velocities of test-particles (Bell 1978b, see also Section 2.3.2 in
Arbutina 2017) which will result in a power-law in momentum: \begin{equation}
N(p) \propto p^{-\mathnormal{\Gamma}}  .
\end{equation}

\noindent Of course, in any case the initial velocity of the particles must be larger than that of the shock, i.e. the particles must be supra-thermal.
Another interesting process is the re-acceleration of CRs which is most commonly assumed to be a diffusive process in which Galactic CRs are re-accelerated through the
second order Fermi process in the ISM (e.g.~Drury \& Strong 2015). However, one can consider the re-acceleration of pre-existing CRs by the shock through DSA mechanism.
In this case the pre-existing CRs are actually seed particles that are injected in DSA process (e.g.~see Section 2.3.4 in Arbutina 2017). Furthermore, Caprioli et al.~(2018) have
shown that the cosmic-ray particles can be effectively reflected and accelerated regardless of the shock inclination via the so-called diffusive shock re-acceleration mechanism.
They concluded that re-accelerated high-energy particles can drive the streaming instability in the shock upstream and produce effective magnetic field amplification.
In that case, the injection of thermal particles can be triggered even at quasi-perpendicular IS shocks. Finally, it is also possible to account for the CR back-reaction and
to discuss how the non-linear effects influence simple DSA theory presented here (e.g.~Drury 1983, Malkov \& Drury 2001; see also Section 2.4 of this review).

\subsection{DSA - macroscopic approach}

Macroscopic  approach to DSA was developed independently by Axford et al.~(1977), Krymsky (1977) and Blandford \& Ostriker (1978)
by considering diffusion-advection equation for CRs. To derive this equation one should start from collisionless plasma kinetic equation in the following suitable form:

\begin{equation}
\frac{\partial f}{\partial t} + \frac{\partial }{\partial \mathbf{x}}\cdot \Big(f \mathbf{v}\Big) + \frac{\partial }{\partial \mathbf{p}}\cdot \Big(f \frac{d\mathbf{p}}{dt}\Big) = 0,
\label{m1}
\end{equation} transform it to a non-inertial wave frame and average over gyration phase to obtain (see Skilling 1971 and Blandford \& Eichler 1987 for more details):\begin{eqnarray}
\Big(1+\frac{\mathbf{v \cdot w}}{c^2}\Big)\frac{\partial f}{\partial t} + (\mathbf{v}+\mathbf{w})\cdot \nabla f - m\gamma \Big( \frac{\partial \mathbf{w}}{\partial t} + \mathbf{v}\cdot \nabla \mathbf{w}\Big)\cdot \frac{\partial f}{\partial \mathbf{p}} &\approx & \nonumber\\
\frac{\partial f}{\partial t} + (\mu v \mathbf{n}+\mathbf{u})\cdot \nabla f - \Big( \frac{1-\mu ^2}{2} \nabla \cdot \mathbf{u} + \frac{3\mu ^2 -1}{2} (\mathbf{n}\cdot \nabla) (\mathbf{n}\cdot \mathbf{u})\Big) p \frac{\partial f}{\partial {p}} &+& \nonumber \\
\frac{1-\mu ^2}{2} \Big( {v } \nabla \cdot \mathbf{n} + \mu \nabla \cdot \mathbf{u} - 3\mu (\mathbf{n}\cdot \nabla) (\mathbf{n}\cdot \mathbf{u})\Big)  \frac{\partial f}{\partial {\mu}}
&= &
\nonumber \\
\frac{\partial}{\partial \mu} \Big( \frac{1-\mu ^2}{2} \nu _{\rm c} \frac{\partial f}{\partial \mu}\Big) &. &
\label{m2}
\end{eqnarray}
In the last equation, after a transformation $\mathbf{v} \rightarrow \mathbf{v} + \mathbf{w} \approx \mathbf{v} + \mathbf{u}$, distribution function $f$ and momentum $\mathbf{p}$ are
measured in the wave frame, while the coordinates $\mathbf{x}$ are still measured in the inertial frame. {The wave frame} in which CR particles (mainly protons) scatter elastically of,
presumably, Alfv\'{e}n waves, has the velocity with respect to the inertial frame $\mathbf{w} = \mathbf{u} + v_\mathrm{A}  \mathbf{n} \approx \mathbf{u}$, where ${u} \ll c$ is the speed
of the background plasma, $v_\mathrm{A}$ is the Alfv\'{e}n speed and $\mathbf{n}$ the unit vector along the local magnetic field. A consequence of the transformation is that there
is now a collisional, diffusion term on the right-hand side of Eq.~(\ref{m2}) due to waves scattering, but because of the energy conservation, the scattering is only in the {pitch angle
($\mu$ is the cosine of the pitch angle)}, with $\nu _{\rm c}$ representing {collision frequency}, related to the scattering of particles in pitch angle
by hydromagnetic waves -- 'collective interactions', occurring through fluctuating electric and magnetic fields.

{The last term of Eq. (31) is the so-called pitch-angle scattering term in the isotropic form. This is the usual assumption that actually means that the pitch-angle scattering coefficient
is directly proportional to $(1-\mu^2)$, in a special case (see Shalchi et al.~2009 for a thorough derivation). Of course, this is not necessarily correct, but just an approximation.}

To proceed further we will assume that collision frequency is large and that we can expand distribution function into series: $f = f_0 + f_1 + f_2 + \cdots$,
where $f_i = \mathcal{O}(\nu _{\rm c} ^{-i})$. Equating terms of the same order in Eq.~(\ref{m2}) gives us:
\begin{equation}
\frac{\partial}{\partial \mu} \Big( \frac{1-\mu ^2}{2} \nu _{\rm c} \frac{\partial f_0}{\partial \mu}\Big) =0,
\label{m3}
\end{equation}
\begin{eqnarray}
\frac{\partial f_0}{\partial t} + (\mu v \mathbf{n}+\mathbf{u})\cdot \nabla f_0 - \Big( \frac{1-\mu ^2}{2} \nabla \cdot \mathbf{u} + \frac{3\mu ^2 -1}{2} (\mathbf{n}\cdot \nabla) (\mathbf{n}\cdot \mathbf{u})\Big) p \frac{\partial f_0}{\partial {p}} &=& \nonumber \\
\frac{\partial}{\partial \mu} \Big( \frac{1-\mu ^2}{2} \nu _{\rm c} \frac{\partial f_1}{\partial \mu}\Big) .& &
\label{m4}
\end{eqnarray}
Eq.~(\ref{m3}) tells us simply that the distribution function is isotropic in zeroth order.
Averaging Eq.~(\ref{m4}) over $\mu$ gives
\begin{equation}
\frac{\partial f_0}{\partial t} + \mathbf{u}\cdot \nabla f_0 -  \frac{1}{3} \nabla \cdot \mathbf{u} \ p \frac{\partial f_0}{\partial {p}} = 0,
\label{m5}
\end{equation}
while those terms that do not average to zero, with the help of Eq.~(\ref{m5}), give \begin{equation}
\mu v \mathbf{n} \cdot \nabla f_0 + \Big( \frac{1-3\mu ^2}{2} \nabla \cdot \mathbf{u} - \frac{1-3\mu ^2}{2} (\mathbf{n}\cdot \nabla) (\mathbf{n}\cdot \mathbf{u})\Big) p \frac{\partial f_0}{\partial {p}} =
\frac{\partial}{\partial \mu} \Big( \frac{1-\mu ^2}{2} \nu _{\rm c} \frac{\partial f_1}{\partial \mu}\Big) ,
\label{m6}
\end{equation}
or when integrated (Skilling 1971) \begin{equation}
\nu _{\rm c} \frac{\partial f_1}{\partial \mu} = - v \mathbf{n} \cdot \nabla f_0 - \mu \Big(  \nabla \cdot \mathbf{u} -  (\mathbf{n}\cdot \nabla) (\mathbf{n}\cdot \mathbf{u})\Big) p \frac{\partial f_0}{\partial {p}} ,
\label{m7}
\end{equation}
i.e.\begin{equation}
f_1 \approx \bar{f}_1 \mu , \ \ \ \bar{f}_1 =  - \frac{v}{\nu _{\rm c}} \mathbf{n} \cdot \nabla f_0 .
\label{m8}
\end{equation}

If we now return to the original equation and average it to obtain expression for $f_0$, more accurate than Eq.~(\ref{m5})
\begin{equation}
\frac{\partial f_0}{\partial t} + \mathbf{u}\cdot \nabla f_0 + \frac{1}{3} v \mathbf{n}\cdot \nabla \bar{f}_1 -  \frac{1}{3} \nabla \cdot \mathbf{u} \ p \frac{\partial f_0}{\partial {p}} + \frac{1}{3} {v }  \bar{f}_1 \nabla \cdot \mathbf{n}\ = 0,
\label{m9}
\end{equation}
in combination with Eq.~(\ref{m8}) we arrive at the so-called diffusion-advection equation
\begin{equation}
\frac{\partial f_0}{\partial t} + \mathbf{u}\cdot \nabla f_0   =  \nabla (D \mathbf{n} (\mathbf{n}\cdot \nabla) {f}_0 )+  \frac{1}{3} \nabla \cdot \mathbf{u} \ p \frac{\partial f_0}{\partial {p}} ,
\label{m10}
\end{equation}
or in the case of one-dimensional flow with $\mathbf{u} \parallel \mathbf{n}$ ({parallel shock}).
\begin{equation}
\frac{\partial f_0}{\partial t} + u \frac{\partial f_0}{\partial x}   =  \frac{\partial }{\partial x} \Big[ D \frac{\partial f_0}{\partial x} \Big]+  \frac{1}{3} \frac{\partial u}{\partial x} \ p \frac{\partial f_0}{\partial {p}} ,
\label{m11}
\end{equation}
{where $D(x,p)=\frac{v^2}{3 \nu _{\rm c}} $ is coefficient of diffusion parallel to the magnetic field lines, in the case of parallel shocks that we consider here.
It has been shown in Caprioli \& Spitkovsky (2014) and Caprioli et al.~(2015) that perpendicular and generally more oblique shocks do not spontaneously accelerate particles efficiently.
The reason for inefficient acceleration at oblique shocks is that it is more difficult for particles to pre-accelerate and reach injection energy necessary to enter into DSA,
which is an increasing function of shock inclination, and the fraction of injected ions drops exponentially for quasi-perpendicular shocks. An exception already mentioned may be the
re-acceleration of seed particles that are already present, e.g.~galactic CRs, which subsequently also trigger the injection (see Caprioli et al.~2018).}

The general solution of Eq. (40), in the stationary case, obtained by double integration over $x$,
is (see e.g.~Drury 1983):\begin{equation}
f_0(x,p)=A(p)+B(p)\exp\left(\int{\frac{u(x')}{D(x',p)}dx'}\right),
\end{equation}
where $A$ and $B$ are arbitrary functions. In the case of one-dimensional fluid flow in the negative direction, encountering shock at $x=0$, {the velocity is}
\begin{equation}
 u(x)  =  \Bigg\{ \begin{array}{@{\extracolsep{-1.0mm}}ll @{}}
 \ u_1, & \ \ \ x<0 \\
 \ u_2,  & \ \ \ x>0 .
 \end{array}
 \label{m13}
\end{equation}
$D(x',p)$ being
finite {when $x$ goes to infinity, implying that in order} for the distribution function to remain finite downstream, it must be
$f_0(x,p)=F(p)$, and we suppose that it tends to some given distribution far upstream  $f_0(x,p) \rightarrow A(p)$, i.e.
\begin{equation}
 f_0 (x,p)  =  \Bigg\{ \begin{array}{@{\extracolsep{-1.0mm}}ll @{}}
 \ A(p)+B(p)\exp\left(\int{\frac{3 u_1}{\lambda v}dx'}\right), & \ \ \ x<0 \\
 \ F(p),  & \ \ \ x\ge 0 .
 \end{array}
 \label{m14}
\end{equation}
where $D(x,p)=\frac{v^2}{3 \nu _{\rm c}} = \frac{\lambda v}{3}$, $\lambda$ being the mean free path. We assumed that the complete distribution function in the diffusion approximation,
can be given as the sum of isotropic part $f_0$ and the anisotropic part proportional to the gradient of $f_0$,
\begin{equation}
f(x,p)=f_0 (x,p) - \mu \lambda \frac{\partial f_0 (x,p)}{\partial x},
\end{equation}
so that
\begin{equation}
 f (x,p)  =  \Bigg\{ \begin{array}{@{\extracolsep{-1.0mm}}ll @{}}
 \ A+B - \mu \frac{3u_1}{v} B, & \ \ \ x=0^- \\
 \ F,  & \ \ \ x= 0^+ .
 \end{array}
 \label{m16}
\end{equation}

In each of the distribution functions, upstream and downstream, $p$ (but not $x$) is measured in to local fluid frame. Even though the distribution function is invariant,
we still need to make transformations of momenta $p \rightarrow p (1-\mu \frac{u}{v})$, and isotropic parts $f_0 \rightarrow f_0 - \mu \frac{u}{v} p \frac{\partial f_0}{\partial p}$
to move to the shock frame:
\begin{equation}
 f (x,p)  =  \Bigg\{ \begin{array}{@{\extracolsep{-1.0mm}}ll @{}}
 \ A+B - \mu \big( \frac{u_1}{v} p (\frac{\partial A}{\partial p} +\frac{\partial B}{\partial p})+\frac{3u_1}{v} B \big), & \ \ \ x=0^- \\
 \ F - \mu  \frac{u_2}{v} p \frac{\partial F}{\partial p} ,  & \ \ \ x= 0^+ .
 \end{array}
 \label{m17}
\end{equation}
Assuming that the distribution is continuous across the shock, we get the matching conditions:
\begin{equation}
A+B = F,
\end{equation}
\begin{equation}
{u_1} p \Big(\frac{\partial A}{\partial p} +\frac{\partial B}{\partial p}\Big)+ {3u_1} B = {u_2} p \frac{\partial F}{\partial p} .
\end{equation}
Eliminating B from the last two equations and reintroducing shock compression {ratio $R = u_1/u_2$}, we obtain
\begin{equation}
(R-1) p \frac{\partial F}{\partial p} = 3R (A-F) ,
\end{equation}
i.e.
\begin{equation}
F(p) = k_1 p ^{-3R/(R-1)} \int _0 ^p  p'^{3R/(R-1) -1} A(p') dp' + k_2 p ^{-3R/(R-1)}.
\end{equation}
If $A=0$, we get the distribution function
\begin{equation}
F(p) \propto p^{-\mathnormal{\Gamma} - 2} ,
\end{equation} i.e.~the number density $N(p) \propto p^{-\mathnormal{\Gamma}}$, $\mathnormal{\Gamma} = \frac{R+2}{R-1}$, same as in the microscopic approach.
But whatever the distribution $A(p)$ far upstream is, provided that it is softer than a power-law spectrum with a slope $\mathnormal{\Gamma}+2$, the downstream spectrum
at high momenta will always asymptotically tend to a power-law $p^{-\mathnormal{\Gamma} - 2}$ (Drury 1983).

\subsection{Non-linear DSA}

In the test-particle approach or linear DSA it is assumed that the pressure of high-energy particles is small, so that their presence does not modify the shock structure.
If this is not the case, then we are talking about CR back-reaction or non-linear DSA ({NLDSA}; see e.g.~Drury 1983, Berezhko \& Ellison 1999, Malkov \& Drury 2001, Blasi 2002a,b).
Schematic view of a modified in comparison to unmodified shock is presented in Fig.~2. Because of the presence of CR particles ahead of the shock, the density, pressure
and velocity gradients will form upstream in the so-called shock precursor. The jump in these quantities is still present at the so-called subshock, but the total compression
is now larger than the compression at the subshock $R_{\mathrm{tot}} > R_{\mathrm{sub}}$.

CRs modify the shock structure, but the shock modification {then} changes the distribution of CR particles, producing the concave-up spectrum.
This can be understood qualitatively by {noting} that lower-energy CRs will only experience the jump at the subshock and have power-law index
$\mathnormal{\Gamma} \approx (R_{\mathrm{sub}}+2)/(R_{\mathrm{sub}}-1)$, while higher-energy particles will sample a broader portion of the {precursor's} velocity
profile and experience larger {compression, thus having flatter effective} power-law index (Berezhko \& Ellison 1999).

\begin{figure}[h]
  \includegraphics[bb=0 0 1339 711,width=120mm,keepaspectratio]{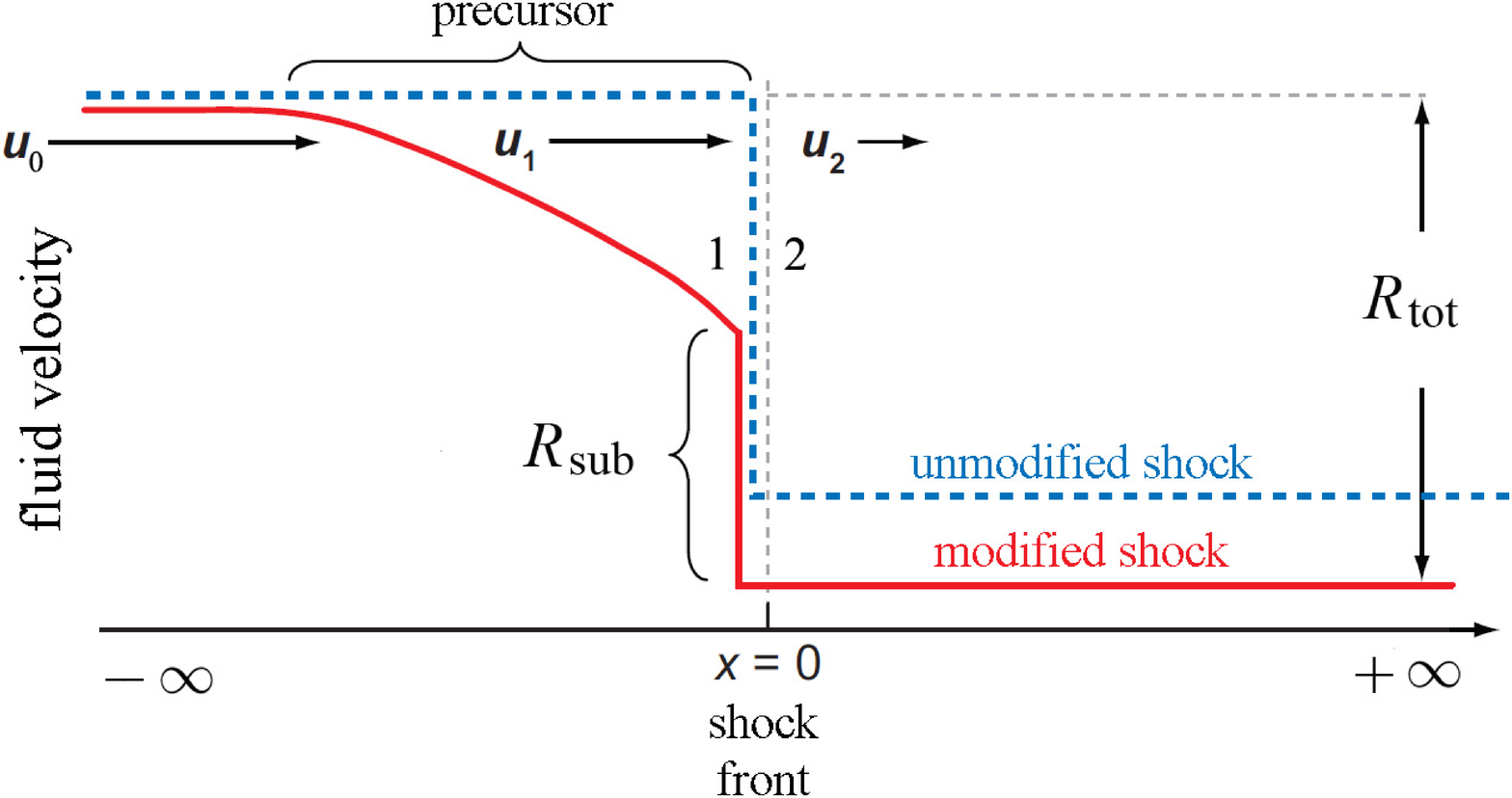}
\caption{Schematic view of a modified shock. Adapted from Reynolds (2008).}
\end{figure}

To show this quantitatively, we will present a semi-analytical model of non-linear DSA given by Blasi (2002a,b) (see also Blasi 2004, Blasi et al.~2005, Blasi et al.~2007,
Amato \& Blasi 2005, {Ferrand 2010} and Pavlovi\'c 2018).We will assume that we measure everything in the shock frame and that the problem is stationary and one-dimensional.
{The} diffusion-advection equation is then
\begin{equation}
\frac{\partial}{\partial x}
\left[ D(x,p)  \frac{\partial}{\partial x} f(x,p) \right] -
u  \frac{\partial f (x,p)}{\partial x} +
\frac{1}{3} \frac{du}{dx}~p~\frac{\partial f(x,p)}{\partial p} + Q(x,p) = 0,
\label{eq:trans}
\end{equation} where $Q(x,p)$ is the so-called injection term, assumed to be in the form $Q(x,p)=Q_0(p) \delta(x)$, where $\delta(x)$ is Dirac's delta function. We saw earlier that momenta
in diffusion-advection equation are actually measured in the fluid frame, but if the particles are energetic enough, our choice of the shock frame will not matter much.
Particles of a certain momentum $p$ will then diffuse upstream ($x<0$) to some distance
\begin{equation}
x_p = \frac{D(p)}{u_p},
\end{equation}
where $u_p$ is some average fluid velocity that will be more precisely defined {later. We implicitly
assume that the diffusion coefficient is an increasing function of momentum (e.g.~linear, $D(p) \propto p$, in the case of Bohm diffusion).
The pressure of CR} particles will slow down the fluid in the precursor, so that its velocity will change from $u_0$ far upstream, to $u_1$ immediately ahead of the subshock
after which it will drop sharply further to $u_2$ downstream.

The first step that we must take is to integrate Eq.~(\ref{eq:trans}) across the subshock (from $x=0^-$ to $x=0^+$, in Fig. 2 these points are marked with 1 and 2, respectively): 
\begin{equation}
\left[ D \frac{\partial f}{\partial x}\right]_2 -
\left[ D \frac{\partial f}{\partial x}\right]_1 +
\frac{1}{3} p \frac{df_0}{dp} (u_2 - u_1) + Q_0(p)= 0,
\end{equation}
where we have assumed continuity of the distribution function $f_2 - f_1 = 0$, i.e. $f_0 = f_1 = f_2$.
By assuming a constant distribution function in downstream region (Blasi 2002a, Reynolds 2008), that is $\left[ D \frac{\partial f}{\partial x}\right]_2=0$,
the last equation {becomes}:
\begin{equation}
\left[ D \frac{\partial f}{\partial x}\right]_1 =
\frac{1}{3} p \frac{df_0}{dp} (u_2 - u_1) + Q_0(p).
\label{eq:boundaryshock}
\end{equation}

The next step is to perform integration of Eq.~(\ref{eq:trans}) again, but now from $x=-\infty$ to $x=0^-$. By using Eq.~(\ref{eq:boundaryshock}) {and applying partial integration
$$\int_{-\infty}^{0^-} u \frac{\partial f}{\partial x} dx
= \left[u f\right]_{-\infty}^{0^-} - \int_{-\infty}^{0^-} f \frac{d u}{d x} dx,$$ we get: \begin{equation}
\frac{1}{3} p \frac{df_0}{dp} (u_2 - u_1) - u_1 f_0 + Q_0(p)+
\int_{-\infty}^{0^-} f \frac{du}{dx}dx +
\frac{1}{3}\int_{-\infty}^{0^-} \frac{du}{d x} p \frac{\partial f}{\partial p}dx = 0.
\label{eq:step}
\end{equation}
}

\noindent We shall now define
\begin{equation}
u_p = u_1 - \frac{1}{f_0} \int_{-\infty}^{0^-} \frac{d u}{dx} f(x,p)dx,
\label{eq:up}
\end{equation}
that represents an average fluid velocity experienced by  particles with momentum $p$ while diffusing upstream. As already mentioned, assuming that $D(p)$ is increasing function of momentum, particles of momentum $p$ will reach only to some $x_p$ and thus sample only a part of the {precursor's} velocity profile.
Hence $u_p$ can be {interpreted physically} as some typical fluid velocity at position $x_p$. The last term on the left-hand side of Eq.~(\ref{eq:step}) {can be transformed to
$$\frac{1}{3}\int_{-\infty}^{0^-} \frac{d u}{d x} p \frac{\partial f}{\partial p} dx = \frac{1}{3}p \frac{d}{dp}\int_{-\infty}^{0^-}f \frac{d u}{d x} dx,$$
so that Eq.~(\ref{eq:step}) becomes:}
\begin{equation}
\frac{1}{3} p \frac{d f_0}{d p} (u_2 - u_p) - f_0 \left(u_p+\frac{1}{3}
p \frac{du_p}{dp} \right) + Q_0(p) = 0 ,
\label{eq:step1}
\end{equation}
where we used
$$
p\frac{d}{dp}\int_{-\infty}^{0^-} \frac{du}{dx} f dx =
p \left(\frac{df_0}{dp} (u_1-u_p) - f_0 \frac{du_p}{dp} \right),
$$
\noindent which follows from the definition of $u_p$. Eq.~(\ref{eq:step1}) represents an ordinary linear differential equation for $f_0(p)$  {when $u_p$ is known and can be integrated to give}:
\begin{eqnarray}
f_0 (p) &=& \int_{p_0}^{p} \frac{d{\bar p}}{{\bar p}}
\frac{3 Q_0({\bar p})}{u_{\bar p} - u_2} \exp\left\{-\int_{\bar p}^p
\frac{dp'}{p'} \frac{3}{u_{p'} - u_2}\left[u_{p'}+\frac{1}{3}p'
\frac{du_{p'}}{d p'}\right]\right\} \nonumber \\
&=& \frac{3 R_{\rm{sub}}}{R_{\rm{sub}}-1} \frac{\eta n_{\rm{1}}}{4\pi p_{\rm{inj}}^3}
\cdot
\exp\left\{-\int_{p_{\rm{inj}}}^p
\frac{dp'}{p'} \frac{3}{u_{p'} - u_2}\left[u_{p'}+\frac{1}{3}p'
\frac{du_{p'}}{d p'}\right]\right\}\!\!.
\label{eq:inje}
\end{eqnarray}
In the above equation we have assumed monochromatic injection of particles with momentum $p_{\rm{inj}}$: $Q_0(p) = \frac{\eta n_{\rm{1}} u_1}{4\pi p_{\rm{inj}}^2} \delta(p-p_{\rm{inj}})$,
where $n_{\rm{1}}$ is gas number density immediately upstream ($x=0^-$) and $\eta$ is the injection efficiency giving the percentage of particles encountering the shock that are injected
into the acceleration process. We have $n_{\rm{1}}=n_0 R_{\rm{tot}}/R_{\rm{sub}} = n_0 R_{\rm{prec}}$, where $n_0$ ia ambient density,  $R_{\rm{sub}}=u_1/u_2$ is the compression at the subshock,
$R_{\rm{tot}}=u_0/u_2$ is the total shock compression and $R_{\rm{prec}}=u_0/u_1=R_{\rm{tot}}/R_{\rm{sub}}$ is the compression in the precursor. While in the case of unmodified strong
adiabatic shock Rankine-Hugoniot jump conditions give $R=4$, here usually $R_{\rm{sub}} <4$ and $R_{\rm{tot}} >4$.

Blasi's model of injection (Blasi et al.~2005) assumes that
\begin{equation}
p_{\rm inj} = \xi p_{\rm th,2},
\label{eq:inj1}
\end{equation}
where thermal momentum  {is} $p_{\rm th,2}=\sqrt{2m_{\rm p}kT_2}$, $T_2$ is downstream temperature and $\xi$ is an injection parameter that can be {related} to injection efficiency,
{from now on designated as $\eta$)}, by requiring continuity of thermal (Maxwell) and non-thermal distribution downstream at $p_{\rm inj}$, that is $f_{\rm th}(p_{\rm inj}) = f_0(p_{\rm inj})$.
From this condition one can find:
\begin{equation}
\eta = \frac{4}{3\sqrt{\pi}}(R_{\rm sub}-1) \xi^3 e^{-\xi^2},
\label{eq:eta}
\end{equation}
where factor $R_{\rm sub}-1$ serves as a kind of regulator -- injection is switched off when $R_{\rm sub} \rightarrow 1$ i.e.~subshock gets smoothed.

If we define dimensionless average fluid velocity $U(p)= U_p = u_p/u_0$, Eq.~(\ref{eq:inje}) {finally becomes}:
\begin{equation}
f_0 (p) = \left(\frac{3 R_{\rm{sub}}}{R_{\rm{tot}} U(p) - 1}\right)
\frac{\eta n_{\rm{1}}}{4\pi p_{\rm{inj}}^3}
\cdot \exp \left\{-\int_{p_{\rm{inj}}}^p
\frac{dp'}{p'} \frac{3R_{\rm{tot}}U(p')}{R_{\rm{tot}} U(p') - 1}\right\}.
\label{eq:inje1}
\end{equation}

\noindent If $U_p\equiv1$, then $R_{\rm{tot}} = R_{\rm{sub}} = R$, and we recover the test-particle solution $f_0 \propto p^{-3R/(R-1)}$. The non-linearity of the problem {lies}
in the fact that $U_p \neq$ const.~and generally $f_0(p)$ will depend on velocity profile $U_p$ through Eq.~(\ref{eq:inje1}), but $U_p$ itself will depend on $f(p)$, in a non-linear fashion.

To find $U_p$ we will use momentum conservation equation that relates quantities far upstream ($x\to-\infty$) with the quantities at $x_p$ {
where} fluid velocity is $u_p$:
\begin{equation}
\rho_0 u_0^2 + P_{\mathrm{th},0} + P_{\mathrm{CR},0} +
P_{\mathrm{w},0} = \rho_p u_p^2 + P_{\mathrm{th},p} + P_{\mathrm{CR},p} + P_{\mathrm{w},p}.
\label{eq:cont}
\end{equation}
In the above equation $\rho$ is the density, $P_{\rm{th}}$ the thermal pressure, $P_{\rm{CR}}$ the non-thermal CR pressure and $P_{\rm{w}}$ the pressure of {MHD waves}.

In the adiabatic approximation
\begin{equation}
\frac{P_{\mathrm{th},p}}{P_{\mathrm{th},0}} =
\left(\frac{\rho_p}{\rho_0}\right)^{\gamma} = \left(\frac{u_0}{u_p}\right)^{\gamma} = U_p^{-\gamma},
\label{eq:adiab}
\end{equation}
where we used mass conservation equation $\rho_0 u_0 = \rho_p u_p$.
In the case of Alfv\'{e}n heating of plasma Berezhko \& Ellison (1999) suggested {a modification to Eq. (64)}:
\begin{equation}
\frac{P_{\mathrm{th},p}}{P_{\mathrm{th},0}} = U_p^{-\gamma}\left[1 + \zeta(\gamma -1)\frac{M_{\mathrm{S},0}^2}{M_{\mathrm{A},0}}(1-U_p^{\gamma})\right],
\label{eq:bif99}
\end{equation}
where $M_{\mathrm{S},0}=\frac{u_0}{c_{\mathrm{S},0}}$ is the Mach's number, $c_{\mathrm{S},0}=\sqrt{\gamma P_{\mathrm{th},0}/ \rho_0}$ ambient sound speed,
$M_{\mathrm{A},0}=u_0/\upsilon_{\mathrm{A},0}$ the Alfv\'{e}n-Mach number, with $\upsilon_{\mathrm{A},0}$ being the Alfv\'{e}n speed.
The Alfv\'{e}n heating parameter $0\leq \zeta \leq 1$ was introduced later by Caprioli et al.~(2009). For $\zeta =0$ we recover Eq.~(\ref{eq:adiab}), i.e.
no Alfv\'{e}n heating, while for $\zeta=1$ there is efficient heating, but then, as we shall soon see, {there is no} magnetic field amplification.

{Let us just note here that one of the most interesting discoveries, which came out of the magnetic field determination using X-ray observations of several young SNRs,
was that the magnetic fields in SNRs were typically $\sim100\ \mathrm{\mu G}$, much larger than might be expected, if they were only caused by the compression of the
IS magnetic field of $\sim5\ \mathrm{\mu G}$ (Vink 2012, Uchiyama et al.~2007). Such high magnetic fields indicate that some kind of a magnetic field amplification
mechanism is operating in these young SNRs. Bell (2004) suggested that magnetic field amplification might be due to a particular plasma instability induced by the streaming
of CR protons away from shocks. Another possibility is to amplify the field as a result of the turbulence induced by the cosmic-ray gradient upstream of the shock acting
on an inhomogeneous ambient medium (so-called Drury instability; Drury \& Downes 2012). The effectiveness of such processes in magnetic field amplification to the
level needed to accelerate CRs up to the PeV domain is still debated in the present literature.}

For the CR pressure in Eq. (\ref{eq:cont}) we {will assume $P_{\rm{CR},0}=0$ (no CR particles far upstream)} and since only the particles with momentum $\geq p$ can reach $x=x_p$, we have:
\begin{equation}
P_{\mathrm{CR},p}= \frac{4\pi}{3} \int_{p}^{p_{\mathrm{max}}} p^3 v(p) f_0(p) dp = \frac{4\pi}{3} \int_{p}^{p_{\mathrm{max}}} \frac{p^4 c^2}{\sqrt{m_\mathrm{p}^2 c^4 + p^2 c^2}} f_0(p) dp,
\label{eq:CR}
\end{equation}
where $v(p)$ is particle speed and $p_{\mathrm{max}}$ is the maximum momentum reached by CR particles which depends on the relevant time-scale of acceleration, escape and losses
(Blasi et al.~2007).

Similarly to CR pressure, we will assume that {MHD waves pressure $P_{\rm{w},0}=0$ far upstream}. In the precursor, closer to the subshock, CR themselves will generate magnetic turbulence
(necessary for their scattering and thus acceleration). This turbulent field will generally have an amplitude larger than the regular field $B_0$ and it can be assumed that this turbulent
magnetic field pressure will be some fraction $\alpha < 1$ of the CR pressure
\begin{equation}
P_{\mathrm{w},p} = \alpha P_{\mathrm{CR},p}.
\label{eq:alfaPw}
\end{equation}
{Based on} the quasi-linear theory  $\alpha \sim \upsilon_{\mathrm{A},0}/u_0$ for the resonant streaming instability (Caprioli et al.~2009), while for the non-resonant {instability}
$\alpha \sim u_0/c$ (Bell 2004). Caprioli et al.~(2009) suggested that
\begin{equation}
\frac{P_{\mathrm{w},p}}{\rho_0 u_0^2} = \frac{1-\zeta}{4 M_{\mathrm{A},0}} U_p^{-3/2} (1-U_p^2),
\label{eq:alfaPw}
\end{equation}
where $U_p^{-3/2}$ is adiabatic compression of the field, and factor $1-\zeta$ account for {the} Alfv\'{e}n heating in Eq.~(\ref{eq:bif99}) -- the wave dumping (and thus the gas
heating) must remain reasonably small for the magnetic field to be substantially amplified ($\zeta < 1$).

Setting $P_{\rm{CR},0}$, $P_{\rm{w},0}$ = 0, in Eq.~(\ref{eq:cont}), dividing by $\rho_0 u_{0}^2$ and inserting Eqs.~(\ref{eq:bif99}), (\ref{eq:CR}) and (\ref{eq:alfaPw}), we obtain:
\begin{eqnarray}
U_p + \frac{U_p^{-\gamma}}{\gamma M_{\mathrm{S},0}^2}\left[1 + \zeta(\gamma -1)\frac{M_{\mathrm{S},0}^2}{M_{\mathrm{A},0}}(1-U_p^{\gamma})\right]  + \frac{4\pi}{3\rho_0 u_0^2} \int_{p}^{p_{\rm max}} dp p^3 v(p) f_0(p) \nonumber \\
 + \frac{1-\zeta}{4 M_{\mathrm{A},0}} U_p^{-3/2} (1-U_p^2) = 1 +
\frac{1}{\gamma M_{\mathrm{S},0}^2}. \ \ \
\label{eq:cont1}
\end{eqnarray}
\noindent Deriving the last equation with respect to $p$ we get, finally:\begin{eqnarray}
\frac{dU_p}{dp}\left\{1 - \frac{U_p^{-(\gamma+1)}}{ M_{\mathrm{S},0}^2}\left[1 + \zeta(\gamma -1)\frac{M_{\mathrm{S},0}^2}{M_{\mathrm{A},0}}\right] - \frac{1-\zeta}{8 M_{\mathrm{A},0}} \frac{U_p^2+3}{U_p^{5/2}}\right\} \nonumber \\
= \frac{4\pi}{3\rho_0 u_0^2}p^3\upsilon(p) f_0(p).
\label{eq:diff_cont}
\end{eqnarray}

Now that we have Eqs.~(\ref{eq:step1}) and (\ref{eq:diff_cont}), we need to know all the parameters appearing in them and the boundary conditions. For fixed Mach and Alfv\'{e}n-Mach numbers
(that is velocity $u_0$ and parameters of the surroundings $\rho_0, P_0, B_0, \gamma$), $\eta$, $\zeta$, $p_{\rm max}$, we must find another relation between $R_{\rm sub},
R_{\rm tot}, R_{\rm prec}$ (knowing that by the definition $R_{\rm tot} = R_{\rm sub}\cdot R_{\rm prec}$), and calculate $p_\mathrm{inj}$. We shall accomplish this by considering jump
conditions at the subshock. Let us start with the momentum conservation equation:
\begin{equation}
\rho_1 u_1^2 + P_{\mathrm{th},1} + P_{\mathrm{CR},1} + P_{\mathrm{w},1}   =  \rho_2 u_2^2 + P_{\mathrm{th},2} + P_{\mathrm{CR},2} +
P_{\mathrm{w},2}.
\label{eq:subshock_cont}
\end{equation}
CR pressure must be continuous across the subshock $P_{\mathrm{CR},1}=P_{\mathrm{CR},2}$, while for the thermal pressure Vainio \& Schlickeiser (1999) derived a modified Rankine-Hugoniot
jump conditions in the presence of plasma's {MHD waves}
\begin{equation}
\frac{P_{\rm th,2}}{P_{\rm th,1}}=\frac{(\gamma+1)R_{\rm sub}-(\gamma-1)\left[1-(R_{\rm sub}-1)\Delta\right]}{(\gamma+1) - (\gamma-1)R_{\rm sub}},
\label{eq:vainio}
\end{equation}
where
\begin{equation}
\Delta = \frac{R_{\rm sub}+1}{R_{\rm sub}-1}\frac{[P_{\rm w}]^2_1}{P_{\rm th,1}}-\frac{2R_{\rm sub}}{R_{\rm sub}-1}\frac{[F_{\rm w}]^2_1}{P_{\rm th,1} u_1},
\label{eq:delta}
\end{equation}
and $[P_{\rm w}]^2_1$, $[F_{\rm w}]^2_1$~are jumps in magnetic field pressure and magnetic energy flux, respectively (we will use notation $[Y]^2_1 = Y_2-Y_1$).
In the case $\Delta=0$, we {obtain} the standard Rankine-Hugoniot jump condition.

Caprioli et al.~(2008, 2009) {found} $[P_{\rm w}]^2_1$ and $[F_{\rm w}]^2_1$ for the {MHD waves}, by considering their transmission and reflection:
$[P_{\rm w}]^2_1=(R_{\rm sub}^2-1)P_{\rm w,1}$, $[F_{\rm w}]^2_1=2(R_{\rm sub}-1)P_{\rm w,1} u_1$, which when inserted in Eq.~(\ref{eq:vainio}) give:
\begin{equation}
\frac{P_{\rm th,2}}{P_{\rm th,1}}=\frac{(\gamma+1)R_{\rm sub}-(\gamma-1)\left[1-(R_{\rm sub}-1)^3\frac{P_{\rm w,1}}{P_{\rm th,1}}\right]}{(\gamma+1) - (\gamma-1)R_{\rm sub}}.
\label{eq:vainio2}
\end{equation}

Eq.~(\ref{eq:subshock_cont}), assuming $P_{\rm CR,1}=P_{\rm CR,2}$, can be transformed to:
\begin{equation}
\frac{\rho_1 u_1^2}{P_{\rm w,1}}\frac{R_{\rm sub}-1}{R_{\rm sub}} + \frac{P_{\rm th,1}}{P_{\rm w,1}}  \left( \frac{P_{\rm th,2}}{P_{\rm th,1}}-1\right) + R_{\rm sub}^2-1=0.
\label{eq:subshock_cont1}
\end{equation}
Let us introduce Mach's number ahead of the subshock $M_{\rm S,1}=u_1/c_{\rm S,1}$, where  $c_{\rm S,1}=\sqrt{\gamma P_{\rm th,1}/\rho_1}$, that can be related to $M_{\rm S,0}$
by using Eq.~(\ref{eq:bif99}): \begin{equation}
\frac{M_{\rm S,1}^2}{M_{\rm S,0}^2}=\frac{\rho_1 u_1^2}{\rho_0 u_0^2} \frac{P_{\rm th,0}}{P_{\rm th,1}}=R_{\rm prec}^{-\gamma-1} \left[1 + \zeta(\gamma -1)\frac{M_{\mathrm{S},0}^2}{M_{\mathrm{A},0}}(1-R_{\rm prec}^{-\gamma})\right]^{-1}.
\label{eq:MS1}
\end{equation}
Finally, from Eqs.~(\ref{eq:vainio2}) and (\ref{eq:subshock_cont1}), after some algebra, we find:
\begin{equation}
M_{\rm S,1}^2 = \frac{2R_{\rm sub}}{(\gamma+1)-(\gamma-1)R_{\rm sub} - 2R_{\rm sub}P_{\rm w,1}^*\left[\gamma - (\gamma-2)R_{\rm sub}\right]},
\label{eq:MS12}
\end{equation}
where we introduced:
\begin{equation}
P_{\rm w,1}^* = \frac{P_{\rm w,1}}{\rho_1 u_1^2} = R_{\rm prec} \frac{P_{\rm w,1}}{\rho_0 u_0^2} = \frac{1-\zeta}{4 M_{\rm A,0}}R_{\rm prec}^{5/2} (1-R_{\rm prec}^{-2}).
\label{eq:Pw1z}
\end{equation}
When we can neglect {MHD waves}, $P_{\rm w,1}\simeq 0$, Eq.~(\ref{eq:MS12}) gives standard {Rankine-Hugoniot} relation:
\begin{equation}
M_{\rm S,1}^2 = \frac{2R_{\rm sub}}{(\gamma+1)-(\gamma-1)R_{\rm sub}}
\Longleftrightarrow R_{\rm sub} = \frac{\gamma+1}{\frac{2}{M_{\rm S,1}^2}+\gamma-1}.
\label{eq:MS12r}
\end{equation}
If $P_{\rm w,1}>0$, for fixed $R_{\rm prec}$, Eq.~(\ref{eq:MS12}) is quadratic in $R_{\rm sub}$:
\begin{equation}
2(\gamma-2)M_{\rm S,1}^2 P_{\rm w,1}^* R_{\rm sub}^2 - \left[2 + (\gamma-1+2\gamma P_{\rm w,1}^*)M_{\rm S,1}^2\right]R_{\rm sub} + M_{\rm S,1}^2(\gamma+1) = 0.
\label{eq:kvad}
\end{equation}
Positive root of this equation will give us $R_{\rm sub}$ as a function of $M_{\rm S,1}$ and $P_{\rm w,1}^*$, and consequently the {total compression $R_{\rm tot}$}.

Finally, we need downstream temperature, in order to calculate $p_\mathrm{inj}$. By using ideal fluid equation of state $P\propto \rho T$ and Eq.~(\ref{eq:bif99}), we have
\begin{equation}
\frac{T_1}{T_0} = \frac{\rho_0}{\rho_1} \frac{P_{\rm th,1}}{P_{\rm th,0}} = R_{\rm prec}^{\gamma-1} \left[1+\zeta(\gamma-1)\frac{M_{\rm S,0}^2}{M_{
\rm A,0}} (1-R_{\rm prec}^{-\gamma})\right],
\label{eq:T1}
\end{equation}
and from Eq.~(\ref{eq:vainio2}) \begin{equation}
\frac{T_2}{T_1} = \frac{\rho_1}{\rho_2} \frac{P_{\rm th,2}}{P_{\rm th,1}}=\frac{(\gamma+1)R_{\rm sub}-(\gamma-1)\left[1-(R_{\rm sub}-1)^3\frac{P_{\rm w,1}}{P_{\rm th,1}}\right]}{\left[(\gamma+1) - (\gamma-1)R_{\rm sub}\right]R_{\rm sub}}.
\label{eq:T2}
\end{equation}

We are now ready for 'shooting for the solution' with an assumed $R_{\rm prec}$ and the initial conditions \begin{equation}
U_{p} (p=p_{\rm inj}) = U_{p} (x=0^{-}) = \frac{u_1}{u_0} = \frac{1}{R_{\mathrm{prec}}},
\label{eq:diff1}
\end{equation} \begin{equation}
\label{eq:limes}
\lim_{p\to p_{\rm{inj}}}f_0 (p) =  \frac{3 R_{\rm{sub}}}{R_{\rm{sub}} - 1}
\frac{\eta n_{\rm{1}}}{4\pi p_{\rm{inj}}^3}.
\end{equation} 
An arbitrarily chosen $R_{\rm prec}$ will, however, not necessarily satisfy the boundary condition \begin{equation}
U_{p} (p=p_{\rm{max}}) = U_{p} (x=-\infty) = \frac{u_0}{u_0} = 1,
\label{eq:diff2}
\end{equation} used to end the integration at $p_{\mathrm{max}}$, so the solution needs to be found iteratively.
To make things simpler, we will introduce dimensionless variables $$
\frac{p}{m_\mathrm{p} c} \rightarrow p,
$$
$$
\frac{4\pi}{3} \frac{m_\mathrm{p}^4 c^5}{\rho _0 u_0^2} f_0 \rightarrow f_0,
$$
and solve simultaneously Eqs.~(\ref{eq:step1}) and (\ref{eq:diff_cont}) in the form \begin{equation}
\frac{1}{3}\Big( \frac{1}{R_{\mathrm{tot}}} - U_p\Big) p \frac{d f_\mathrm{0}}{d p} - \Big( U_p + \frac{1}{3} p \frac{d U_p}{d p} \Big) f_\mathrm{0}= 0 ,
\end{equation}
\begin{eqnarray}
\frac{d U_p}{d p} \Bigg\{ 1 - \frac{U_p^{-(\gamma +1)}}{M_{\mathrm{S},0}^2} \Big[ 1 + \zeta (\gamma -1) \frac{M_{\mathrm{S},0}^2}{M_{\mathrm{A},0}} \Big] - \frac{1 -\zeta}{8 M_{\mathrm{A},0}} \frac{U_p^2 +3}{U_p^{5/2}} \Bigg\}
=  \frac{p^4 f_\mathrm{0}}{\sqrt{1 + p^2}} .
\end{eqnarray}

\begin{figure}[h!]
  \includegraphics[bb=0 0 792 633,width=58mm,keepaspectratio]{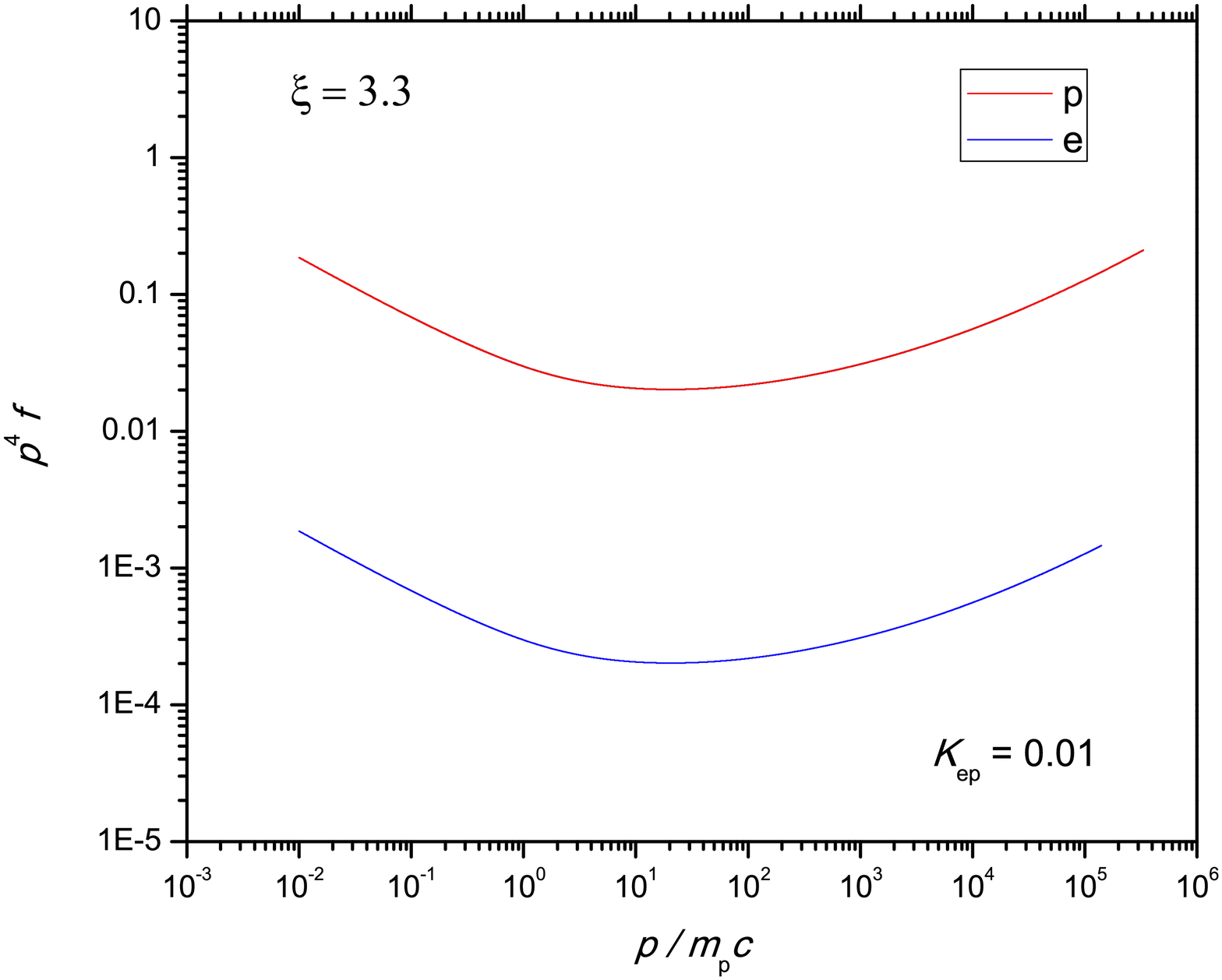}
  \includegraphics[bb=0 0 792 633,width=58mm,keepaspectratio]{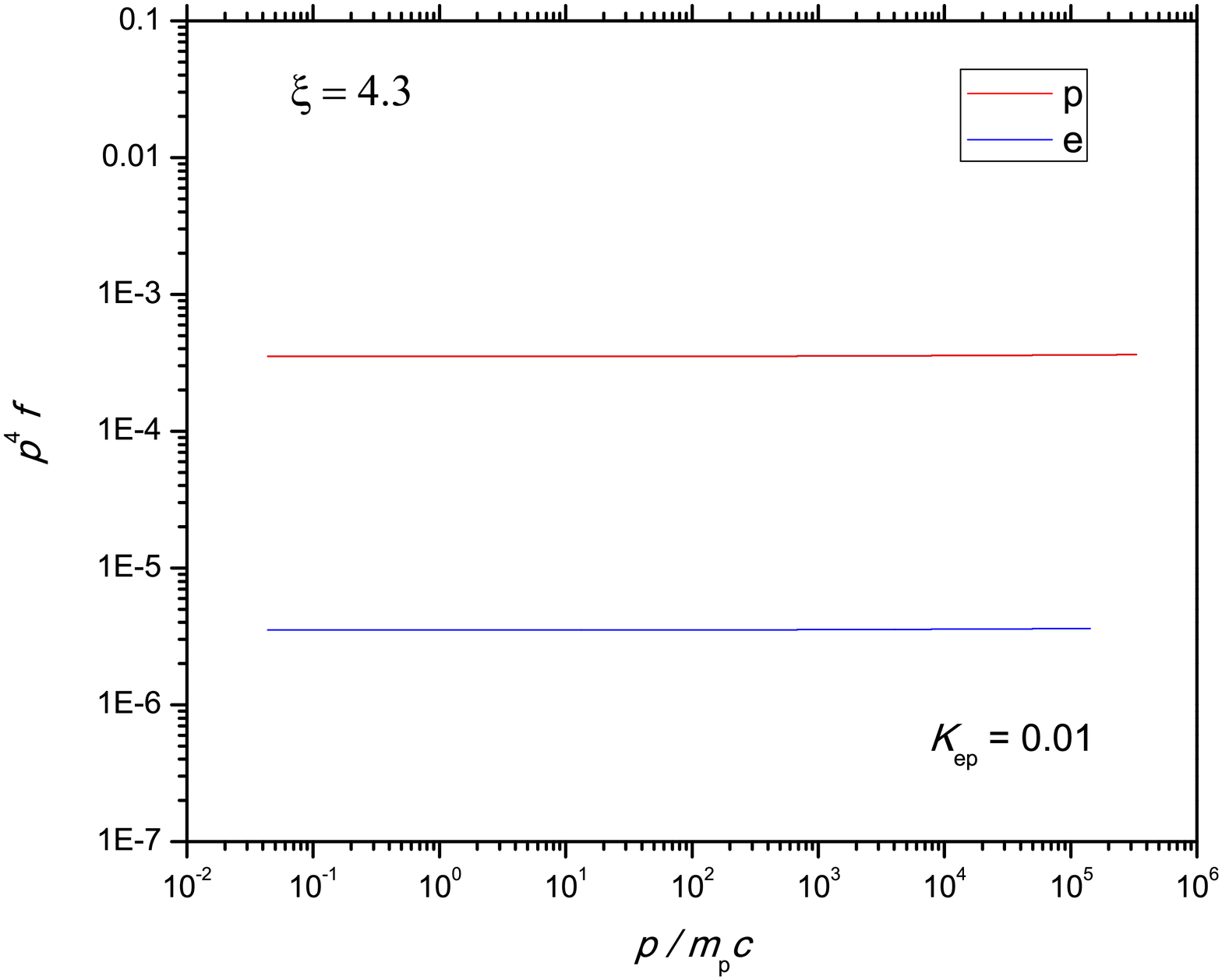}
\caption{Proton and electron spectra for injection parameter $\xi =3.3$ (left) and $\xi =4.3$ (right).}
\vspace{2mm}
\end{figure}

In Fig.~3 we give the results for two cases: strongly-modified shock ($\xi=3.3$) and a test-particle case ($\xi =4.3$), as in Caprioli et al.~(2010).
Other parameters in two cases are the same: shock velocity $u_0$ = 5000 km/s, ambient density $n_\mathrm{H} \sim 0.1 \mathrm{cm}^{-3}$, temperature $T_0 = 10^4$ K,
magnetic field $B_0$ = 5.3775 $\mu$G, equal Mach and Alfv\'{e}n-Mach numbers of 135 and Alfv\'{e}n heating parameter $\zeta=0.5$. For the case $\xi=3.3$, subshock
and the total compressions are $R_\mathrm{sub}=3.081$, $R_\mathrm{tot}=10.236$, $p_\mathrm{inj}=0.01$ and $p_\mathrm{max} = 3.316 \cdot 10^5$, while for $\xi$ = 4.3,
$R_\mathrm{sub}=3.999$, $R_\mathrm{tot}=4.018$, $p_\mathrm{inj}=0.0437$, $p_\mathrm{max}$ is the same. For 'practical' purposes, e.g.~modeling the radio synchrotron
emission of astrophysical sources such as supernova remnants (e.g.~Berezhko \& V\"olk 2004, Pavlovi\'c 2017, Pavlovi\'c et.~al 2018), besides proton spectrum it
is also crucial to know electron spectrum. It is usually assumed that the proton and electron spectra are parallel and that $p_\mathrm{inj}$ for protons and electrons are the same
(although how this is accomplished for electrons is still uncertain) in order for both CR species to cross and recross the subshock (with an assumed thickness
$\approx r_{\rm g} \propto p_\mathrm{th}$ of protons) unaffected. The only unknown parameter left is then the electron-to-proton number ratio at high energies $K_\mathrm{ep}$, for which
in Figs.~3 and 4 we assumed $K_\mathrm{ep}=1:100$, in accordance with the observed ratio for Galactic cosmic-rays. The extreme cases presented in Fig.~3 depict
a general behavior -- shocks with low $\xi\lesssim 3.5$ have high injection efficiency $\eta$ and are highly modified, producing a typical concave-up spectrum; shocks with high
$\xi\gtrsim 4$ have low injection efficiency and will produce basically a test-particle spectrum $f_0 \propto p^{-\mathnormal{\Gamma}-2}$ with $\mathnormal{\Gamma} \approx 2$.
{It is important to note at the end that these models do not account for losses and particle escape, since it is assumed that $P_{\rm{CR},0}=0$ far upstream (for alternative see e.g. Caprioli et al. 2010).}

%
%
%
%
%


\section{Observational signatures of particle acceleration in ISM: a quick overview}

Instead of the summary, we present here a quick overview of the most important observational signatures of particle acceleration in ISM.

As already seen in the previous sections, DSA can be held responsible for the production of the non-thermal ensemble of cosmic-ray particles which in the simplest
test-particle case has a power-law energy distribution. Actually, the particle energy spectral index, derived from the theory seems to be in very good accordance with the
present observations of primary CRs of non-solar origin in the vicinity of planet Earth ($\mathnormal{\Gamma}\approx2.7$ up to $\sim10^{15}\ \mathrm{eV}$).

Furthermore, the presence of ultra-relativistic charged particles moving in the external global magnetic field will generally cause significant production of
synchrotron radiation. The history of radio astronomy teaches us that one of the first detected objects that glow in the radio sky are indeed SNRs, in particular
Cas A remnant (Reber 1944). In fact, the non-thermal radio-continuum spectra of SNRs, shaped by the synchrotron emission, unambiguously pointed to the presence of high-energy charged particles linked
to the IS shock waves. Strictly speaking, however, this only provided the evidence for electron acceleration, whereas the observed CR spectrum in the Earth's
neighborhood consists mainly of protons and other heavy ions. Of course, for massive charges synchrotron radiation is indeed emitted, but at a much lower efficiency.
As we noted earlier, protons and heavier particles can be accelerated very efficiently to $\sim10^{15}\ \mathrm{eV}$. Electrons can also be accelerated to the ultra-relativistic
energies ($\sim10^{12}\ \mathrm{eV}$) by DSA mechanism and this lower maximal energy of electrons comes from the very rapid energy losses induced by the inverse Compton, non-thermal bremsstrahlung and synchrotron radiation.

The energy spectrum of relativistic electrons which is in the form of a power-law in test-particle DSA is transformed in the power-law
radio continuum spectrum. The particle energy index from the energy spectrum of CRs can be transformed in the so-called spectral index of the radio spectrum $\alpha$,
by the simple linear relation $\mathnormal{\Gamma}=2\alpha+1$. A radio continuum may then be characterized by the power-law form $S_{\nu}\propto\nu^{-\alpha}$,
where $S_{\nu}$ is the radio flux density at frequency $\nu$. It is easy to conclude that the value for spectral index derived directly from test-particle DSA theory is 0.5.
As we know from the observational data, the large majority of radio spectra of Galactic SNRs have $\alpha=0.5$ (more generally between 0.2 and 0.8; see Green 2017
as well as Uro\v sevi\' c 2014, for more details). This really seems to be an excellent confirmation of the theoretical predictions. Furthermore, for a standard value
of the mean Galactic magnetic field strength ($\sim5\ \mu\mathrm{G}$), GeV electrons are responsible for the synchrotron emission at higher radio-frequencies,
and TeV electrons for X-rays.

{X-ray synchrotron emission that is linked to the SNRs is usually associated with the existence of a pulsar wind nebula (or plerion). Still, over the last several decades
the evidence for electron acceleration in SNRs has been significantly enhanced by the detection of X-ray synchrotron radiation from the shells of several young remnants (Koyama et al.~1995).
This X-ray synchrotron emission implies that CR electrons are indeed accelerated up to energies of around 10 to 100 TeV. Furthermore, such young SNRs generally have an X-ray synchrotron spectra with rather
steep indices $\mathnormal{\Gamma}=2-3.5$ which points to a steep underlying electron energy distribution. Such a detected steepness, on the other hand,
indicates that the synchrotron X-ray emission is in fact caused by the electrons close to the maximum energy of the CRs electron energy distribution.}


Furthermore, $\gamma$-ray observations are a promising way to study CR acceleration in the SNRs, especially in the TeV and even PeV range (e.g.~Gaggero et al.~2018).
We note here that the $\gamma$-rays may not necessarily come from the pulsars and pulsar wind nebulae, but from the SNR shells, too. The Cherenkov $\gamma$-ray telescopes
(like H.E.S.S., MAGIC and VERITAS) have detected around 16 TeV sources that are associated with SNRs, while few tens of firm identifications at GeV energies
are present in the first Fermi catalogue of SNRs (Acero et al.~2016). The great importance of $\gamma$-ray astronomy is reflected in the fact that it can give us a direct view of the CR nuclei,
accelerated via DSA. These cosmic-ray protons and heavy ions produce $\gamma$-ray emission if they collide with the ambient IS atomic nuclei. As a result, among the other products of
such a collision process, neutral pions are created, which then decay in two $\gamma$-ray photons (the so-called hadronic scenario). This allow us to trace high-energy CRs above GeV energy
(see Inoue 2019, and references therein). However, two other important $\gamma$-ray radiation processes originate from CR electrons (the so-called leptonic scenario).
Interactions with background photons result in the inverse Compton up-scattering, whereas interactions with ions in the SNR result in bremsstrahlung radiation. Usually, inverse Compton scattering
is a more dominant leptonic process in young SNRs. However, it is generally very difficult to distinguish between hadronic and leptonic origin of $\gamma$-rays
(Sano et al.~2019). The advanced TeV and other $\gamma$-ray telescopes like the Cherenkov Telescope Array could in principle help us to infer more on the role of
SNRs in the efficient CR production.

{In the case of very young SNRs, i.e.~strong collisonless shocks, we expect that the effects of the non-linear DSA can cause a slightly concave up synchrotron spectrum
(Reynolds \& Ellison 1992, Jones et al.~2003, de Looze et al.~2017). In fact, from Section 2.4, we have learned that
a non-linear DSA theory predicts that the particle energy spectrum steepens at low energies and also flattens at higher energies. This immediately leads
to the curved synchrotron (radio to microwave continuum) spectrum, that can be crudely modeled by a pure power-law with a varying spectral index, such
as in the case of famous SNR Cas A (Oni\'c \& Uro\v sevi\'c 2015). Of course, further observations of the integrated radio up to microwave continuum of SNRs is of great
importance as any possible deviations from the known theory can give us new clues about physics of the observed emission. We need reliable flux density
estimates at as many as possible different continuum frequencies. However, this is connected with serious observational problems, such as transparency
issues regarding the Earth's atmosphere. The new confirmations of the theoretically predicted radio spectral features at high radio frequencies could be
expected from future observations by for instance, ALMA (Atacama Large Millimeter/submillimeter Array) telescope. As a final note, another consequence of the spectral curvature
is that the actual X-ray synchrotron brightness cannot be just simply estimated from an extrapolation of the radio synchrotron spectrum -- it should be brighter
than the one expected (Vink et al.~2006, Allen et al.~2008).}

{Finally, a word of caution regarding the non-linear DSA theory presented earlier in this paper. Several authors claim that the concavity of the spectrum contradicts present $\gamma$-ray
observations (see e.g.~Caprioli 2011). One way of removing the concavity was found in Ferrand et al.~(2014). It was shown that at perpendicular shocks,
the concavity can be removed by replacing the Bohm diffusion coefficient by a more realistic form.

Another important question is related to the observational fact that several young pre-Sedov SNRs exhibit rather steep (but not curved) radio spectral indices ($\alpha>0.5$).
Neglecting the non-linear hydrodynamic effects due to the CR pressure in the precursor, Bell et al.~(2011) demonstrated that the oblique-shock effects can produce a steep spectrum
if the high velocity shock of a young SNR has a tendency towards a quasi-perpendicular configuration. They noted that either a magnetic field amplification in the precursor due
to CR streaming or expansion into a circumstellar wind supporting a Parker spiral may produce a quasi-perpendicular shock geometry. Bell et al.~(2011) also concluded that
the Galactic CR spectrum is most probably formed by a complex mixture of non-linear, oblique-shock and momentum-dependent CR escape effects. In the case of large shock inclinations,
acceleration efficiency decreases and one would expect steeper spectrum. However, acceleration efficiency for shock inclinations greater than $60^{\circ}$
is found to be decreasing, reaching practically zero percent (see Figure 3 in Caprioli \& Spitkovsky 2014) and quenching particle acceleration (except in the case of diffusive shock
re-acceleration mechanism). Spectral steepening in young SNRs has also been discussed by Bell et al.~(2019). They proposed a new process in which the loss of CR energy to turbulence and magnetic
field during the CR acceleration at the non-relativistic quasi-parallel shocks is responsible for the spectral steepening. One should bear in mind that such a process is indeed a non-linear
effect in the sense that it depends on the shock velocity and on non-linear turbulent amplification of magnetic field, but it does not depend on the ratio of the CR pressure to the
kinetic pressure at the shock. On the other hand, using particular numerical simulation that incorporates three-dimensional hydrodynamic modeling and non-linear kinetic theory of CR
acceleration in parallel shocks, taking into account the non-linear back reaction of accelerated particles on the fluid structure, Pavlovi\'c (2017) found that the steep (but not significantly
curved) radio spectral index of the youngest known SNR G1.9+0.3 can be solely explained by means of the efficient NLDSA. The most probable scenario is the one that incorporates
more than one of the proposed processes that act at the same time.}

In addition, a significant contribution of the second order Fermi (or stochastic) acceleration mechanism is usually proposed to shape the radio-continuum of several evolutionary old Galactic
SNRs with spectral indices $\alpha$ less than 0.5 (Schlickeiser \& F\"{u}rst 1989, Ostrowski 1999). However, contribution of the secondary electrons left over from the decay of charged
pions (if an SNR is interacting with a molecular cloud), or just a simple thermal contamination, or even the intrinsic thermal bremsstrahlung radiation from the SNRs, etc., can also
cause such flat spectral indices as observed in some, usually evolutionary old SNRs in high density environment (Uchiyama et al.~2010, Oni\'c 2013).

\begin{acknowledgements}
{We thank the anonymous referee for useful comments and suggestions that greatly improved the quality of this paper.}
We acknowledge the financial support of the Ministry of Education, Science, and Technological Development of the Republic of Serbia through the project No.~176005
"Emission Nebulae: Structure and Evolution". The authors thank Dragana Momic for careful reading and correction of the
manuscript.
\end{acknowledgements}





\end{document}